\documentclass[sigconf]{acmart}

\usepackage[normalem]{ulem}

\AtBeginDocument{%
  }

\copyrightyear{2026}
\acmYear{2026}
\setcopyright{cc}
\setcctype{by}
\acmConference[CHI '26]{Proceedings of the 2026 CHI Conference on Human Factors in Computing Systems}{April 13--17, 2026}{Barcelona, Spain}
\acmBooktitle{Proceedings of the 2026 CHI Conference on Human Factors in Computing Systems (CHI '26), April 13--17, 2026, Barcelona, Spain}
\acmDOI{10.1145/3772318.3790498}
\acmISBN{979-8-4007-2278-3/2026/04}

\begin{document}

\title{From Efficiency to Meaning: Adolescents’ Envisioned Role of AI in Health Management}

\author{Jamie Lee}
\affiliation{%
  \institution{Informatics}
  \institution{University of California, Irvine}
  \city{Irvine}
  \state{California}
  \country{USA}
}

\author{Kyuha Jung}
\affiliation{%
  \institution{Informatics}
  \institution{University of California, Irvine}
  \city{Irvine}
  \state{California}
  \country{USA}
}

\author{Cecilia Lee}
\affiliation{%
  \institution{School of Medicine}
  \institution{University of California, Irvine}
  \city{Orange}
  \state{California}
  \country{United States}}

\author{Lauren MacDonnell}
\affiliation{%
  \institution{School of Medicine}
  \institution{University of California, Irvine}
  \city{Orange}
  \state{California}
  \country{United States}}

\author{Jessica Kim}
\affiliation{%
  \institution{School of Medicine}
  \institution{University of California, Irvine}
  \city{Orange}
  \state{California}
  \country{United States}}

\author{Daniel Otterson}
\affiliation{%
  \institution{School of Medicine}
  \institution{University of California, Irvine}
  \city{Orange}
  \state{California}
  \country{United States}}

\author{Erin Gregg Newman}
\affiliation{%
  \institution{School of Medicine}
  \institution{University of California, Irvine}
  \city{Orange}
  \state{California}
  \country{United States}}

\author{Emilie Chow}
\affiliation{%
  \institution{School of Medicine}
  \institution{University of California, Irvine}
  \city{Orange}
  \state{California}
  \country{United States}}

\author{Yunan Chen}
\affiliation{%
 \institution{Informatics}
 \institution{University of California, Irvine}
 \city{Irvine}
 \state{California}
 \country{USA}}
 
\renewcommand{\shortauthors}{Lee et al.}

\begin{abstract}
While prior research has focused on providers, caregivers, and adult patients, little is known about adolescents’ perceptions of AI in health learning and management. Utilizing design fiction and co-design methods, we conducted seven workshops with 23 adolescents (aged 14-17) to understand how they anticipate using health AI in the context of a family celiac diagnosis. Our findings reveal that adolescents have four main envisioned roles of health AI: enhancing health understanding and help-seeking, reducing cognitive burden, supporting family health management, and providing guidance while respecting their autonomy. We also identified nuanced trust and a divided view toward emotional support from health AI. These findings suggest that adolescents perceive AI's value as a tool that moves them from efficiency to meaning–one that creates time for valued activities. We discuss opportunities for future health AI systems to be designed to encourage adolescent autonomy and reflection, while also supporting meaningful, dialectical activities. 
\end{abstract}




\maketitle

\section{INTRODUCTION}
The rapid evolution of artificial intelligence (AI) has accelerated the integration of AI-based technologies across diverse areas of healthcare and everyday health experiences \cite{jiang_artificial_2017, schwalbe_artificial_2020, davergne_wearable_2021, lee_human-ai_2021, milne-ives_effectiveness_2020}. Ranging from clinical care and diagnostics \cite{asan_artificial_2020, wang_brilliant_2021, cai_human-centered_2019, fogliato_who_2022, hu_systematic_2025, jacobs_designing_2021} to self-management of health conditions and improving wellness \cite{stromel_narrating_2024, fitzpatrick_delivering_2017, you_user_2022, van_arum_selective_2025}, various \textit{health AI systems} are being designed to support health-related decisions and behaviors of clinicians and laypersons. However, the majority of prior work in this field has concentrated on AI use in clinical settings.

HCI research on health AI for personal health management---defined as AI integrated, consumer-facing applications that help individuals monitor, interpret, and manage their health in daily settings---has been growing but remains underexamined. As AI becomes increasingly implemented into products like into mobile health apps \cite{bhatt_emerging_2022, deniz-garcia_quality_2023, ryan_users_2025, su_analyzing_2020}, symptom checkers \cite{you_medical_2021, you_user_2022}, self-tracking systems \cite{stromel_narrating_2024, rashik_ai-enabled_2025}, or mental health chatbots \cite{inkster_empathy-driven_2018, sweeney_can_2021, abreu_enhancing_2024, milne-ives_effectiveness_2020}, these systems are aiming to better support people with health-related skills and management activities \cite{abreu_enhancing_2024, rashik_ai-enabled_2025, stawarz_co-designing_2023, sweeney_can_2021}. Health AI applications can also include generative AI chatbots (e.g., ChatGPT) as these LLM-based tools are already being used for health tasks, such as information-seeking, symptom triage, self-tracking, and addressing mental well-being needs \cite{al_shboul_investigating_2024, jung_ive_2025, ayre_new_2024, yun_online_2025}. By integrating AI-powered functionalities (e.g., predictive analytics, personalized content curation, and natural language interaction \cite{van_arum_selective_2025, sweeney_can_2021, xiao_powering_2023, stromel_narrating_2024}), many technologies are demonstrating the ability to improve our day-to-day health practices. 

Despite the promise of health AI, existing studies consistently highlight critical issues that impact how people perceive, engage with, and adopt these systems \cite{detjen_who_2025, yoo_patient_2024, rashik_ai-enabled_2025, kim_how_2024, young_patient_2021, richardson_patient_2021, tucci_factors_2022, asan_artificial_2020}. Trust, misinformation, bias, and privacy issues are several of the many concerns voiced by healthcare providers, patients, and the general public \cite{obermeyer_dissecting_2019, seyyed-kalantari_underdiagnosis_2021, crawford_excavating_2021, lee_deepfakes_2024}, which can further affect people’s perceptions and usage of health AI. Furthermore, there is ongoing debate about the role of AI in providing emotional support. HCI studies have shown that empathy and human-like features in AI are often valued for their potential in providing emotional support \cite{you_beyond_2023}. However, excessive reliance on AI for this role can result in harmful emotional dependence \cite{laestadius_too_2024, zhang_dark_2025} or feelings of unease and discomfort \cite{kim_can_2018, lee_understanding_2025}. 

While prior HCI literature has primarily focused on adult users, such as clinicians \cite{detjen_who_2025, wang_brilliant_2021, guo_evaluating_2025, cai_human-centered_2019}, parents \cite{petsolari_socio-technical_2024, ramgopal_parental_2023}, and adult patients \cite{yoo_patient_2024, rashik_ai-enabled_2025, kim_how_2024}, adolescents remain an understudied yet crucial population. Adolescence is a critical window of brain and socio-emotional development, often called “a period of opportunity,” where young individuals begin to develop lifelong skills and behaviors \cite{bonnie_promise_2019, dekovic_risk_1999}. During this stage, adolescents experience increasing desire for autonomy and risk-taking as they mature into adulthood, all of which shape how they make decisions and learn how to manage health. A recent report from Common Sense has shown adolescents’ growing usage of AI, particularly for school and general information \cite{madden_dawn_2024}. However, little is known about how they perceive or would use AI in the context of health. Given their curiosity, risk-taking tendencies, and greater use of AI, adolescents are at a very consequential moment in their lives, and \textbf{it is imperative that we investigate their perceptions of AI to ensure that we are designing effective, developmentally appropriate technologies}. Without these insights, misunderstandings \cite{kennedy_public_2023, bewersdorff_myths_2023}, misuse \cite{tracy_united_states_responsible_2024, lee_deepfakes_2024}, and excessive trust \cite{kapania_because_2022} of health AI technologies by this population can lead to serious negative consequences. Taken together, we aim to explore the potential of health AI in health learning and management for adolescents and sought to answer these questions: 
\begin{description}
    \item \textit{RQ1: How do adolescents envision the role of AI in supporting their individual and family health management?}
    \item \textit{RQ2: What concerns do adolescents perceive in the health AI systems they envision?}
\end{description}

To address these questions, we ground our work in HCI’s established traditions of design fiction and co-design to explore how adolescents envision using health AI for learning and managing health. These methods are commonly utilized to investigate people’s perceptions of particular topics while deepening our understanding of the present and futures they imagine \cite{fitton_co-designing_2018, aarts_snoozy_2022, rubegni_dont_2022, chowdhury_co-designing_2023, lee_understanding_2025}. We collaborated with a team of expert physicians to develop a series of progressive scenarios in the context of a family health condition: celiac disease. Then, using the scenarios and design activities, we held seven workshops with 23 adolescents (aged 14-17).

We found that participants expressed qualified optimism towards health AI. Their expectations of health AI included enhancing health understanding, easing cognitive burden, mediating family collaborative health, and offering guidance while respecting autonomy. Collectively, our participants positioned health AI as a promising tool to enhance their learning and autonomy, where it could support their transition to independence. They also saw AI’s ability to be efficient as a way to save time for meaningful activities (e.g., family time), rather than as a replacement for human experiences. Their perceived concerns about trust were affected by factors such as differing health contexts and misconceptions about AI. Together, these findings suggest that adolescents have a critical view toward health AI: while they value AI’s efficiency, they are divided on receiving direct emotional support from it. We contend that participants perceived AI as enabling a shift from efficiency to meaningful engagement by off-loading routine tasks to create space for intrinsically valuable activities. They prefer AI to act as a supplementary guide that augments, rather than leads or replaces, human-led activities. Drawing from our findings, we reflect on these insights and provide design implications for future health AI technologies for adolescents. 

Our work makes the following contributions:
\begin{enumerate}
    \item Through design fiction and co-design workshops, we provide a rich account of adolescents’ envisioned roles of AI for health management, illustrating how AI can serve as an engaging tool for fostering health learning and autonomy.
    \item We advance understanding of adolescents’ critical stance on health AI–perceiving AI as a supportive, human-centered tool that guides rather than replaces human-led activities.
    \item We offer nuanced insights into adolescents' perceived concerns, particularly around trust and emotional support from health AI, highlighting how these perceptions are contextually shaped.
\end{enumerate}

\section{RELATED WORK}

\subsection{Health AI for Personal Health Management}
Health AI for personal health management holds great promise for helping individuals understand and improve their personal health and well-being across contexts, including fitness \cite{van_arum_selective_2025, stromel_narrating_2024}, eating habits \cite{james_chronic_2023}, and mental well-being \cite{sweeney_can_2021, wong_voice_2024, fitzpatrick_delivering_2017}. For instance, mHealth apps and fitness trackers can improve physical fitness by providing personalized plans, tracking progress, and offering context-based notifications \cite{stromel_narrating_2024, chen_investigating_2024}. AI symptom checkers can help users interpret their symptoms and suggest plausible diagnoses, with recent studies indicating that individuals valued an empathetic, conversational style from these systems \cite{you_user_2022, you_beyond_2023}. Studies have also demonstrated that AI chatbots designed for mental health can help alleviate symptoms \cite{sweeney_can_2021, fitzpatrick_delivering_2017, inkster_empathy-driven_2018}. Prior work on large language model (LLM)-based chatbots has highlighted their widespread implementation and ability to provide timely and tailored information, potentially fostering a deeper understanding of users’ health status \cite{al_shboul_investigating_2024, ayre_new_2024}. The growing functionalities of health AI are enabling individuals to manage their health and well-being independently through tracking their health status, finding reliable information, and assessing their symptoms.

As healthcare becomes more patient-centered, health AI can stand to support individuals, especially those with chronic conditions, across more situations. Such systems can enable patient monitoring, deliver personalized health information, and strengthen self-management skills. For example, many patients can seek health information using LLM-based chatbots. Xiao et al.’s team collaborated with experts worldwide and developed an AI chatbot designed to supply credible and easily accessible information about COVID-19 to individuals \cite{xiao_powering_2023}. They found that users obtained more accurate answers with the AI chatbot compared to a search engine, indicating the prospect of AI to enhance health information-seeking experiences. However, as with any technology, complications arise, which will be discussed in the next section.

\subsection{Issues Regarding Health AI}
Despite optimism about the capabilities and benefits of health AI, significant challenges remain \cite{mohsin_khan_towards_2025, seyyed-kalantari_underdiagnosis_2021, koenecke_racial_2020, obermeyer_dissecting_2019, tucci_factors_2022, lee_deepfakes_2024}. One significant concern is bias, in which factors, such as data, algorithms, and even the developers, can lead AI systems to produce inequitable outcomes with potentially serious negative consequences \cite{seyyed-kalantari_underdiagnosis_2021, koenecke_racial_2020, obermeyer_dissecting_2019}. For example, in dermatology, an AI-powered diagnostics tool demonstrated significantly higher accuracy for individuals with White skin types at 69.9\% compared to those with Black skin types at 17\% \cite{kamulegeya_using_2023}. Another major issue is hallucinations, or generated information that seems to make sense but is factually incorrect or fabricated \cite{bhattacharyya_high_2023, agarwal_medhalu_2024}. Hallucinations stem from large language model (LLM) applications, such as ChatGPT, and as more and more people are using such tools for health-related inquiries, the risks associated with hallucinations have increased. For example, Agarwal et al. found that individuals without medical training failed to detect hallucinations in LLM-generated health content \cite{agarwal_medhalu_2024}. Finally, another issue surrounding health AI is the possible drawbacks of receiving emotional support from AI. While recent work on AI conversational agents and chatbots has shown the growing benefits of a non-judgmental space \cite{jung_ive_2025, lee_i_2020, kim_can_2018, ma_evaluating_2024}, empathetic responses \cite{fitzpatrick_delivering_2017, daher_empathic_2020, you_beyond_2023}, and stigma-free emotional support \cite{lopatovska_capturing_2022, ma_evaluating_2024, you_beyond_2023}, the risks have also been highlighted \cite{ma_evaluating_2024, jung_ive_2025, sweeney_can_2021, haque_overview_2023, bae_brandtzaeg_when_2021}. Receiving emotional support from AI may foster excessive trust and reliance on these systems as a primary source of help rather than seeking professional care \cite{bae_brandtzaeg_when_2021, jung_ive_2025, haque_overview_2023}. For instance, Brandtzaeg et al. evaluated young people’s perception of social support from chatbots and found that some participants reported trusting chatbots more than humans, despite expressing privacy concerns \cite{bae_brandtzaeg_when_2021}. Given that most previous research has focused on the adult population, we aim to extend the literature by examining how adolescents perceive the risks of health AI, in the context of their distinct developmental needs.

\subsection{Adolescent Health Management in HCI}
HCI scholars have explored how technologies can support adolescents’ health and well-being in various ways. A primary focus has been supporting the development of health-related skills, such as information seeking, data management, and decision-making \cite{oh_hey_2025, cha_collaborative_2025, rahman_adolescentbot_2021, su_creating_2024, freeman_putting_2023, zehrung_investigating_2021}. For example, prior work has shown that adolescents frequently turn to social media and the internet for health information, with their own standards of trusting the sources \cite{freeman_how_2023, freeman_how_2018, freeman_role_2020}. Researchers have explored other technologies, such as chatbots, to provide more accurate health information with Rahman et al. finding that adolescents valued privacy protection when discussing health issues \cite{rahman_adolescentbot_2021}. Beyond information seeking, HCI research has extensively examined the considerable effort adolescents undertake in handling their chronic illness and how technologies can support them, highlighting the promise of digital tools for learning and self-management \cite{raj_my_2019, zehrung_transitioning_2024, cha_collaborative_2025, cha_transitioning_2022, hong_care_2016, su_creating_2024, su_data-driven_2024, silva_unpacking_2023, ankrah_me_2023}. For instance, Zehrung et al. found that adolescents with various chronic illnesses shared similar desires to manage routine health tasks independently and utilized online resources when needed, suggesting the need to design technology that not only augments their autonomy but also ensures their safety \cite{zehrung_transitioning_2024}. Additionally, adolescents’ health is directly shaped by their family structure as parents typically handle major responsibilities, including health management \cite{bonnie_promise_2019}. HCI work has examined family informatics and collaborative management for both well-being and chronic illness, showing that technology can have beneficial effects. For example, as children become older, adolescence marks a period when they become more actively involved in their health, often co-managing with their parents \cite{lee_mobile_2023, su_data-driven_2024, su_creating_2024, hong_care_2016}.

Despite the availability of digital tools and support from families, adolescents face barriers in effectively managing their health for several reasons. First, the transition from pediatric to adult care remains a challenge \cite{gray_barriers_2018, zehrung_transitioning_2024, ankrah_plan_2023, james_chronic_2023}. Adolescents often have limited knowledge about their health, disease, and the transition process, reflecting broader gaps in health literacy \cite{gray_barriers_2018}. While they are motivated to seek information, they lack the strategies to properly assess health information found online \cite{freeman_how_2018, freeman_how_2023}, despite being regarded as “digital natives.” Second, many adolescents struggle with self-management skills, partly influenced by their parents \cite{yu_conflicts_2023, zehrung_transitioning_2024, kaziunas_caring_2017}. Tensions arise as adolescents’ desire for autonomy and independence increase, while parents struggle with giving up control as they are concerned about their health and safety \cite{hong_care_2016, zehrung_transitioning_2024}. Studies have also revealed tensions that arise within family collaborative management, particularly around sharing data. While many adolescents were comfortable sharing much of their data, they resisted full disclosure, suggesting conflict between their emerging independence and privacy \cite{pina_personal_2017, zehrung_transitioning_2024, lee_understanding_2025}.

Although HCI scholars have long investigated adolescents’ interactions with health technologies, this population remains understudied in the context of emerging AI systems. The rapid advancement of AI makes this gap increasingly pressing, particularly as health AI holds not only great potential to transform health but also significant capacity to cause harm. Building on these bodies of work, our study contributes insights into how adolescents envision AI’s role in supporting their health learning and management.

\section{METHODS}
To investigate adolescents’ perceptions of health AI, we conducted seven workshops with 23 adolescents (aged 14-17). We employed design fiction and co-design methods to elicit reflective engagement with health AI technologies. Our study was approved by the university’s Institutional Review Board (IRB). In this section, we describe the following: our study design, procedure, recruitment, data analysis process, and limitations.

\subsection{Study Design}
Our study design combined two complementary methodologies: design fiction and co-design. Design fiction uses fictional scenarios, narratives, or prototypes to explore speculative futures and elicit open-ended responses from the participants \cite{linehan_alternate_2014, blythe_research_2014, sterling_cover_2009, dunne_speculative_2013}. These scenarios are commonly used as probes to help participants imagine a specific situation and to evoke responses about the technology, its potential benefits, risks, and other ethical considerations \cite{rubegni_dont_2022, wong_real-fictional_2017, collyer-hoar_its_2024}. Co-design has been widely used with the youth to actively engage participants in discussion and research \cite{collyer-hoar_its_2024, fitton_co-designing_2018, freeman_putting_2023, yip_laughing_2019, mcnally_co-designing_2018, aarts_snoozy_2022}. Through creative activities and group discussions, researchers can gain deeper insights into the needs and behaviors of younger audiences, informing the development of technologies that are appropriately tailored to children and adolescents \cite{aarts_snoozy_2022, chowdhury_co-designing_2023, freeman_putting_2023}.

Based on these methods, our workshop was designed collaboratively with several domain experts, including one senior HCI professor and four medical doctors who specialize in clinical pediatric care. We worked together to create the scenarios, select the right health condition, and plan the design activities. First, we selected a family-based scenario as adolescents under 17 years old typically live at home with their families. Chronic health conditions are highly prevalent in families, with an estimated 76.4\% of U.S. adults experiencing at least one chronic condition \cite{watson_trends_2025}. Additionally, we intentionally had adolescents design for others, as previous research suggests that this approach can foster perspective-taking and empathy, helping participants move beyond their own experiences to consider diverse users’ needs and contexts \cite{van_mechelen_developing_2018, lo_noel_2025}.

After several discussions with our team of expert doctors, we decided on celiac disease for several reasons. First, we chose a physical health condition to clarify the scope for this study, leaving mental health for a future study. Second, we selected a serious but not life-threatening condition that was relatively unfamiliar to our target audience. Around 1.4\% of people around the world have celiac disease \cite{beyondceliac_celiac_2025, singh_global_2018}, signaling its prevalence. Finally, given that celiac disease has no medical cure and must be managed through lifelong gluten avoidance, we chose it intentionally to require adolescents to consider day-to-day health management rather than medication-based or procedural treatments.

We focus on adolescents because they are in a unique yet critical developmental phase \cite{bonnie_promise_2019}. At ages 14-17, they assume more responsibilities, including their health, seek independence, and develop habits that will affect their adulthood. They also have access to digital technologies such as mobile phones and, therefore, AI. As they develop these skills with almost unlimited access to AI, it is essential that adolescents have the correct understanding, knowledge, and skills to effectively utilize health AI in their daily lives.

\subsection{Workshop Procedure}
Workshops were conducted over Zoom with activities carried out on Miro, a digital whiteboard. Each lasted about 120 minutes with a break after the second activity. Per IRB protocol, sessions were recorded with consent. After a brief overview of the research goal, introduction, icebreaker, and overview of health AI, the lead researcher guided three scenarios-based activities, reconvening for group debriefs after each. Sessions ended with concluding questions and open discussion. We describe the scenarios and activities in detail below. 

\subsubsection{Scenario \& Activity 1: Initial Discovery \& Health Information Seeking}
In the first activity, we explored what adolescents initially would ask about a new health condition and how they would use AI to obtain information. To start, participants were asked to imagine that their mother had been recently diagnosed with celiac disease and to write down the initial questions that came to their minds. Then, they were prompted to use ChatGPT, a popular commercial LLM chatbot, to ask their questions and to reflect on their experience and the responses. The first activity was designed to (1) introduce celiac disease as a context, (2) engage participants in using AI for health-related tasks (e.g., searching for information), and (3) prompt them to imagine using AI for health management purposes.

\subsubsection{Scenario \& Activity 2: Brainstorming \& Co-Designing AI Solutions for Supporting Mother's Celiac Disease Management}
For the second scenario, adolescents were asked to imagine how AI could support their mothers in managing celiac disease. They were prompted to first think about problems that their mothers might encounter after being diagnosed. Together, participants brainstormed problem statements and discussed the most prominent challenges their mothers and families might face. After, they selected one to design an AI solution for and were asked: \textit{"Imagine we are in the future, year 2035, what do you think AI can do by then to help solve these problems?"} They were given template images of a smartphone, smartwatch, computer, robot, speakers, and VR headset, and prompted to draw, sketch, or write out their ideas in Miro. We decided to provide templates after the first workshop as it was difficult for participants to draw in Miro with time constraints. The goal of the second activity was to explore how adolescents envision AI supporting health management within their family.

\subsubsection{Scenario \& Activity 3: Storyboarding Feelings of Personal Worry \& Symptom Investigation}
For the third scenario, participants were prompted to imagine experiencing celiac-like symptoms (e.g., stomach ache, indigestion, and feeling sick) and thinking they might have celiac disease. For the activity, they were asked to draw and write out how AI would support them in moments of worrying about getting celiac disease, investigating symptoms, and discussing with a trusted adult. The objective of the last activity was to shift the health context from adolescents' mothers to themselves and understand their views on how AI could support navigating a personal potential diagnosis.

\renewcommand{\arraystretch}{1.2}
\begin{table}[h!]
\centering
\begin{tabular}{p{2cm} p{2cm} p{1.5cm} p{1cm}}
\toprule
\textbf{Workshop \#, Participant \#} & \textbf{Pseudonym} & \textbf{Gender} & \textbf{Age} \\
\midrule
W1, P1 & Brian & Male & 15 \\
W1, P2 & Daisy & Female & 16 \\
W1, P3 & Lily & Female & 15 \\
W2, P4 & John & Male & 16 \\
W2, P5 & Michael & Male & 17 \\
W2, P6 & Sam & Male & 16 \\
W2, P7 & Jack & Male & 17 \\
W3, P8 & Rae & Female & 17 \\
W3, P9 & Sarah & Female & 15 \\
W3, P10 & Ashlynn & Female & 17 \\
W4, P11 & Jenny & Female & 17 \\
W4, P12 & Brianna & Female & 16 \\
W5, P13 & Samuel & Male & 16 \\
W5, P14 & Jason & Male & 17 \\
W5, P15 & Nick & Male & 14 \\
W6, P16 & Ivy & Female & 15 \\
W6, P17 & Susie & Female & 17 \\
W6, P18 & Serene & Female & 14 \\
W6, P19 & Emma & Female & 16 \\
W7, P20 & Ian & Male & 16 \\
W7, P21 & Ruby & Female & 15 \\
W7, P22 & Julia & Female & 16 \\
W7, P23 & Irvin & Male & 17 \\

\bottomrule
\end{tabular}
\caption{Participant Information}
\label{tab:table1}
\end{table}

\subsection{Recruitment and Participants} After obtaining IRB approval for the procedure above, we recruited 23 adolescents (13 girls, 10 boys) who were 14-17 years old (mean age: 16, SD: 0.95) from several high school summer programs that were held at an academic institution in the western United States. The programs focused on teaching a variety of topics, such as healthcare careers and computer science, to high school students. The first author went and spoke about the research project during the programs and handed out recruitment material after. We also utilized snowball and convenience sampling methods to recruit. After interested participants filled out the screening questionnaire, eligible participants and their parents were contacted via email to obtain assent through DocuSign. After obtaining assent and consent from both parties, participants were scheduled for a workshop based on availability. 23 adolescents completed this process and attended workshops from July-August 2025. 

\subsection{Data Analysis}
After each workshop, the first author cleaned the transcripts and began analysis. There was a total of 769 minutes of video data (not including the introduction and icebreaker at the beginning). The first author started with qualitative memos \cite{birks_memoing_2008} after each workshop to document initial thoughts and reflections and shared the memos with the research team. Memos captured early reflections on adolescents' expectations of AI when navigating a new health condition (e.g., providing simplified information and actionable next steps) as well as mixed reactions to receiving emotional support from AI. After the final workshop, the first author began coding the transcripts inductively on Atlas.ti. We followed Braun and Clarke's thematic analysis approach which involved progressing from data familiarization to open coding to the construction and refinement of the final themes \cite{braun_using_2006, clarke_thematic_2017}. Our initial codes included “\textit{cope with stress,}” “\textit{provides knowledge,}” “\textit{maintain quality of life,}” and “\textit{checking symptoms.}” While the full research team met biweekly to share updates, the first and last authors met weekly to review, refine, and organize the codes into broader themes. After rounds of deliberation, we iteratively finalized the overall themes, identifying adolescents’ expectations of AI in health learning and management.

\subsection{Limitations}
This study has several limitations. Participants were recruited from the Western United States, which may limit generalizability and introduce selection bias. The mean age of our participants was 16 years old, meaning our findings may reflect the views of older adolescents (aged 15-17) rather than younger ones (aged 12-14), whose views may differ \cite{nguyen_examining_2022}. We held single-session workshops rather than continuous, multi-day workshops. Although continuous workshops offer more time and space for engagement and reflection \cite{kumar_co-designing_2018, chowdhury_co-designing_2023}, single day workshops improved recruitment. We also did not directly collect data about participants' AI literacy levels and familiarity with AI tools, which may limit our ability to interpret how prior experience shapes their views, trust, and concerns. Finally, we did not include parents in the workshops, as we aimed to focus on adolescents and their unique perspectives of AI for health management. Future research should include various perspectives of family members as parents often oversee their children’s health \cite{bonnie_promise_2019, su_creating_2024} whereas adolescents’ greater technological fluency may shape distinct viewpoints and dynamics around health AI.

\section{FINDINGS}
In this section, we present our findings. Synthesizing across the three scenario-based activities, we identified four primary expectations adolescents had of AI in health management: enhancing health understanding and help-seeking, reducing cognitive burden, mediating families and collaborative health management, and providing guidance while respecting adolescents’ autonomy (RQ1). We also found that perceived concerns centered on trust and expectations of emotional support from AI, which were highly context-dependent and varied with the stakes of the situation and the type of task (RQ2). In presenting these findings, we refer to the participants by their pseudonym and workshop identifier (e.g., Brian (W1, P1) denotes Workshop 1, Participant 1).

\subsection{Adolescents' Envisioned Roles of AI in Health Learning and Management (RQ1)}
\subsubsection{AI for Enhancing Health Understanding and Help-Seeking} 
Throughout the activities, we found that adolescents primarily considered AI as a tool for learning and understanding a new chronic condition for both someone else or themselves. Participants highlighted AI's value in providing introductory information about a new health condition, in supporting self-assessment of symptoms, and in preparing difficult conversations with a trusted adult.

\begin{figure*}[!h]
  \centering
  \includegraphics[width=15.5cm]{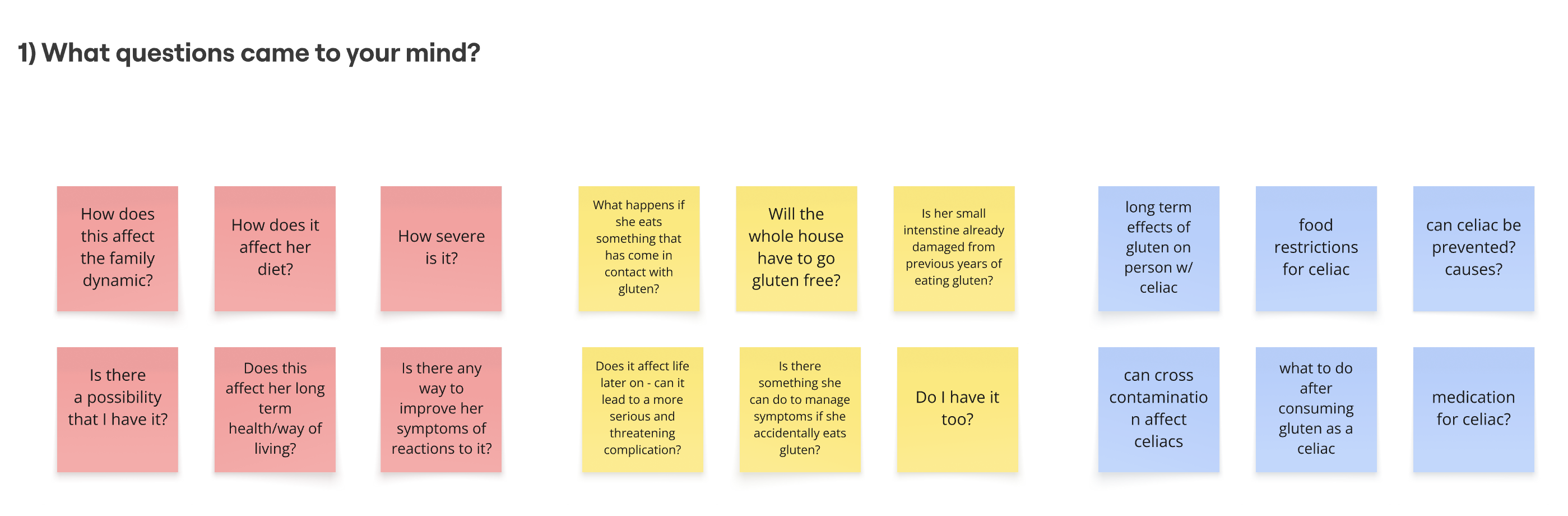}
  \caption{Examples of initial questions raised by adolescents when confronted with mother's celiac disease diagnosis}
  \label{tab:figure1}
  \Description{The image shows two rows of post-it notes with questions.}
\end{figure*}

\textbf{Providing preliminary information.} When confronted with a (hypothetical) life-changing diagnosis of their mothers, adolescents voiced AI's usefulness as a source of preliminary information. In the beginning, they raised two types of questions: disease-specific and family life. Most questions focused on understanding the condition–how it works and long-term consequences, as seen in \textbf{Figure 1}. For example, Samuel (W5, P16) wanted to know \textit{“why does the body reject gluten?”} Many probed about the severity of celiac disease, such as \textit{“is it fatal?”} and \textit{“does this condition worsen with age?”} The second set of questions was focused on how their families’ lives would change. Ray (W3, P8) considered how \textit{“it would affect the family dynamic… how do we eat around the mom?”} As celiac disease is a food-related health condition, families’ meals and cooking routines could change as Jenny’s (W4, P11) initial question was \textit{“What will we eat?”} After asking ChatGPT these questions, many participants reported feeling \textit{"informed"}. As Michael noted, \textit{"[ChatGPT] told me a lot of things I didn't know about celiac disease, so I feel informed."} Similarly, Sarah (W3, P9) described herself using AI as \textit{"an information box that has a lot of information based off the whole Internet."} Adolescents noted that AI could be beneficial for specifically obtaining groundwork information when first learning about a topic, such as a new health condition.

When brainstorming about supporting their mothers’ celiac disease, one of the main challenges that adolescents brought up was their own limited understanding of the condition. They worried that their knowledge gaps could unintentionally worsen symptoms and make daily routines more burdensome. For example, it was crucial to have information about which foods have gluten or not, especially in situations outside the home, such as restaurants or social settings. Nick (W5, P15) noted, \textit{“if you don't have a good understanding of what foods you can't eat, your problem could get worse if you keep eating foods with gluten. [...] This is important because you don't want to damage your intestine even more.”} They expressed apprehension about not knowing what steps to take after an accidental exposure to gluten, noting that the family would need to be prepared before (by preventing cross-contamination) and after exposure (by knowing what to do next or having medication in hand). To solve this particular issue, many adolescents proposed AI solutions that included gluten-checking as the primary feature, with several expanding the capability to include meal suggestions and instructions for handling accidental gluten consumption (\textbf{see Figure 2}). Participants designed AI apps (W1, W2, W5, W7) and a portable food scanner (W4), which were centered on providing information about gluten in the food. Participants from Workshop 3 created a combined robot-smartwatch AI system, where the robot checked food for gluten, and the smartwatch delivered emergency protocols after gluten exposure. Sarah (W3, P9) suggested that the smartwatch could \textit{"give recommendations about if you aren't feeling good, what to do. [...] It can give you more information about [your condition]."} This indicates that our participants felt that AI could provide quick and accessible information, especially during health-related contingencies.

\textbf{Assisting in self-assessment of symptoms.} When the context shifted from supporting their mothers to imagining their own symptoms, adolescents pictured AI as an assistant for analyzing and understanding what they were experiencing. Ray (W3, P8) explained that \textit{“the first thing I would do is ask [AI] about the symptoms that I'm experiencing or ask about how I'm feeling or why I might be feeling that I have the disease.”} Participants viewed AI not only as a source of information but also as a tool to help them make sense of their symptoms and overall health conditions. As an example, Sam (W2, P6) highlighted AI’s capabilities as a diagnostic tool: \textit{“Because I’m giving [AI] more personal info and more specific recounts of [symptoms], I think it can tell me if it’s more of an allergy versus if it’s actually celiac.”} Nick (W5, P15) similarly noted AI could point to non-celiac causes, saying, \textit{“[AI] can help you discover and find out what's causing your symptoms because it might not be celiac disease, it could totally be something else.”} They imagined AI as a tool in their investigation to verify their suspected health condition. While most participants did not fully trust AI to completely diagnose them, they trusted AI enough to seek preliminary information and better understand their symptoms.

\textbf{Preparing and initiating discussions with a trusted adult.} Adolescents highlighted AI's prospective role in helping them organize their thoughts and prepare for initiating conversations about their symptoms and worries with a trusted adult. Many participants anticipated AI to be supportive by \textit{“suggesting a way to approach the conversation in a really productive way”} (Ray, W3, W8) or even \textit{“help tell me what to say”} (Sarah, W3, P9). Sarah explained, \textit{“I think sometimes in stressful situations, we don’t go to people because we don’t know how to approach them or what to say.”} These comments highlight the difficulties of adolescents in initiating serious discussions with adults. Beyond initiating discussions, adolescents also desired support in preparing and organizing their thoughts before their conversations. Sam (W2, P6) expected AI to help \textit{“organize my thoughts before I present all this info.”} For Ian (W7, P20), he wanted AI to assist him with figuring out how he could articulate his needs more clearly so that he \textit{“doesn’t sound awkward or that I’m emotionally unstable. [...] [To] talk to [an adult] in a way in which they can understand me.”}

In contrast, there were a handful of adolescents who believed they did not need AI in this situation. For instance, Michael (W2, P5) questioned AI’s role in a conversation with a trusted adult, stating, \textit{“I wasn’t sure how to implement AI to that because you're just talking to a trusted adult, which is a doctor giving you statistics about yourself.”} Jenny (W4, P11) also noted, \textit{“I don’t think AI really needed to be used in that scenario because we could tell our mom. [...] She would be able to understand and help us.”} She further explained that \textit{“since it’s just human interaction, there is no need for AI.”} These reflections suggest that some adolescents struggled to understand how AI might be relevant when they were interacting with a trusted expert or a close adult and preferred to handle the interpersonal interaction themselves.

\subsubsection{AI for Reducing Cognitive Load}
Adjusting to celiac disease would make life more complicated, as managing a new health condition requires mental effort for learning, decision-making, and adapting daily habits. Adolescents discussed using AI as a means to reduce cognitive burden by helping with time demands and information overload. 

\begin{figure*}[!h]
  \centering
  \includegraphics[width=10cm]{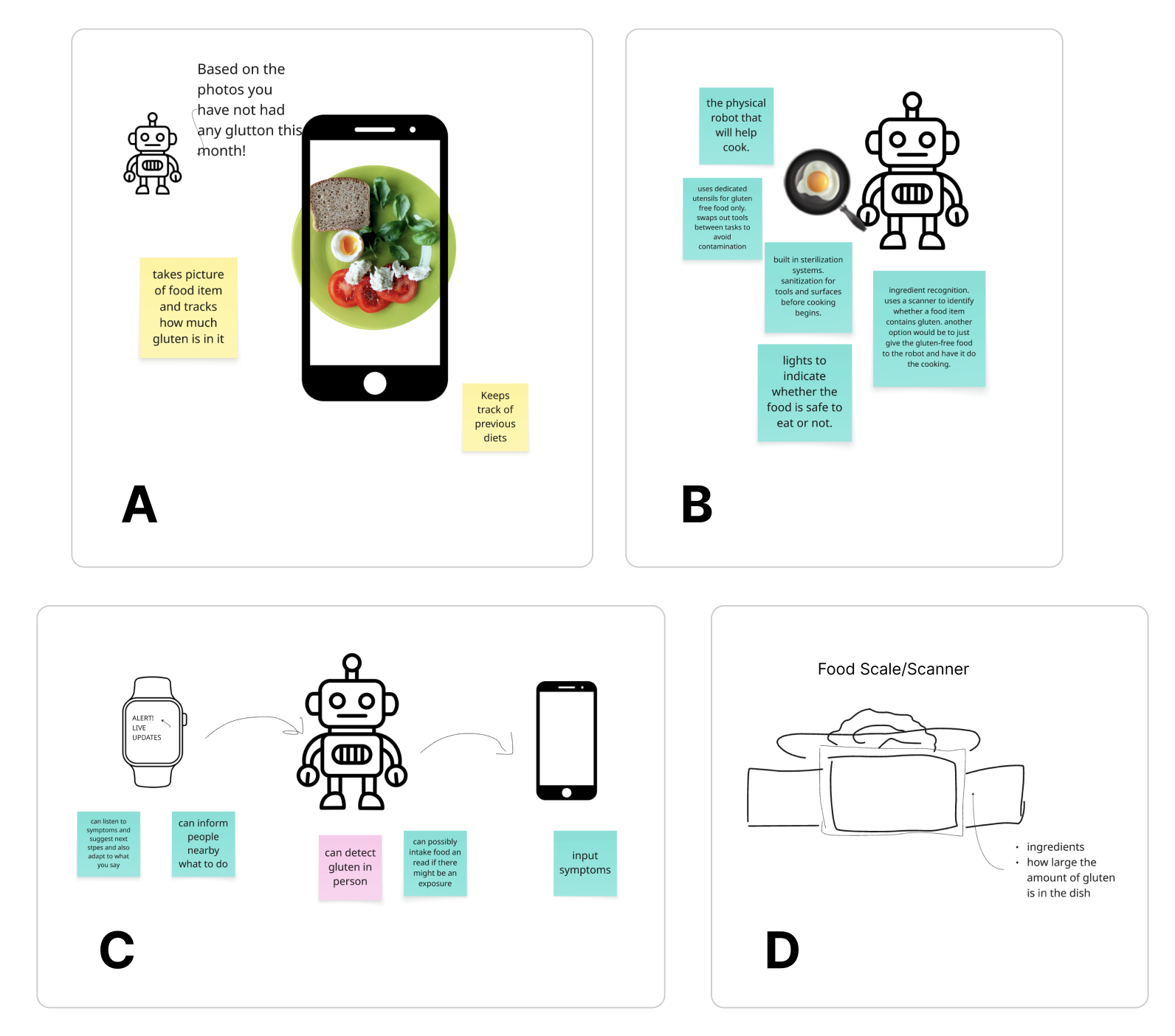}
  \caption{Examples of co-designed health AI systems for mother's celiac disease management (A: an AI app by Workshop 7 participants, B: an AI robot by Workshop 6 participants, C: an AI system with a robot and a smartwatch by Workshop 3 participants, and D: an AI food scale scanner by Workshop 4 participants)}
  \label{tab:figure2}
  \Description{The image shows two rows of post-it notes with questions.}
\end{figure*}

\textbf{Saving time and increasing efficiency.} 
Adolescents acknowledged that handling a new chronic disease could make everyday routine tasks much more difficult, which contributed to their belief that AI could help save time and reduce cognitive burden. Given mothers' central role in household management, they emphasized how responsibilities, like cooking, cleaning, and grocery shopping, could take more time and effort due to the new reality of avoiding gluten. Ivy (W6, P16) worried that \textit{“[celiac disease] will disrupt the daily lives of our family because my mom does cook most of the time.”} Similarly, Ashlynn (W3, P10) was concerned about sterilization and cross-contamination in their homes because \textit{“moms do a lot of the cleaning and dishes and just, and they’re often in the kitchen.”} Family social outings could become more complicated as Brian (W1, P1) commented,\textit{“You can't just go anywhere because a lot of restaurants have a lot of options with gluten in them, and you want to just be safe, but at the same time still being able to enjoy while you eat.”} Based on their concerns, adolescents believed that AI could assist in these moments, as shown in their proposed AI solutions. For instance, Michael (W2, P5) described how their proposed AI app would help his mother grocery shop: \textit{“she can’t really already read English like that good, so she could have trouble reading ingredients.”} He asserted that instead, \textit{“you could be doing some other things that are productive instead of frustrating and put more effort into something that would take a robot 10 minutes to complete.”} Michael (W2, P5) pictured AI assisting his mother in finding gluten-free groceries more quickly by presenting information in a manner that she could quickly comprehend. Participants felt that AI had the capabilities to process information easily and complete mundane tasks that would become difficult because of celiac disease. Beyond daily tasks at home, adolescents commented that their families enjoyed going out to eat, but searching for gluten-free options could also be time-consuming and mentally taxing. Brian (W1, P1) noted, \textit{“Rather than having to stop by each restaurant, ask them [about gluten], and do all that, that wastes a lot of time. This [AI app] would save us a lot of time.”}

\textbf{Simplifying complex and overwhelming health information.}
In addition, adolescents viewed AI as playing an important role in helping prepare for a life-changing diagnosis---for themselves or for a family member---by providing quick, clear, and comprehensive responses to their questions. Reflecting on their experience using AI (e.g., ChatGPT) to search for health information, participants appreciated that the health information was presented clearly, utilizing tactics such as summaries, bulleted lists, bolded keywords, and even emojis. Michael (W2, P5) asked for a summary and \textit{“it did a really good job at it because it literally made a graph telling me what it is and what it means. It just made it really helpful for me.”} New health information can be overwhelming, and participants found that AI’s structured, well-formatted response made complex topics easier to digest. For instance, Irvin (W7, P23) shared, \textit{“It was formatted in a way that I can read it pretty well–not a lot of jargon and a lot of just helpful information.”} Similarly, Sarah (W3, P9) highlighted AI’s powerful capability to \textit{“simplify really complicated things, and so it can give you information and links from every corner of the Internet that could take hours for you to find yourself”}. It could even benefit younger audiences as Lily (W1, P3) expressed: \textit{“[ChatGPT] bolded the keywords which is really nice to make it easy to skim and kind of concise in that sense. But it seems like anyone could really comprehend what it’s saying, which could be helpful if you’re younger and trying to figure out something with health.”} Overall, participants valued AI’s ability to take copious amounts of information and transform it into content that felt approachable, concise, and developmentally appropriate.

\subsubsection{AI for Mediating Family Life and Collaborative Health Management}
Although adolescents were prompted to think about their mothers' diagnosis, they acknowledged that their families' lives would be affected as well. When designing AI solutions for their mothers, adolescents reflected on how the condition could create strain and disruption in their family life and identified opportunities to become more actively involved. They conceptualized AI as a mediator for collaborative family health management by creating more family time together, sharing health data within their families, and directly involving adolescents in health AI for managing family health.

\textbf{Creating more family time.} 
Our participants believed that health AI could afford opportunities to spend more time together as families. Building on AI's envisioned role in reducing cognitive burden, adolescents expected AI to take over daily tasks and create time for meaningful activities with families. The proposed AI solutions were all situated in their families' day-to-day routines, and many focused on supporting or taking over certain tasks. \textit{“With more time, family time could also increase… especially if a parent is the one that's cooking a lot in the household, then that parent doesn't get to be with the family during that time,”} Susie (W6, P17) noted, a sentiment many shared. Looking into the future, Jack (W2, P7) reflected that \textit{“if AI takes care of a lot of the tasks that are time consuming… [that time] could be instead focused on family time.”} The proposed health AI concepts would directly support family bonding time–such as cooking together–as participants included features that provided gluten-free recipes. John (W2, P4) described that their AI app would \textit{“create menu items and it could help for family bonding, like cooking together as a family,”} creating avenues for adolescent involvement. 

\textbf{Sharing health data among the family.} 
Most adolescents preferred health AI to support data sharing among family members, facilitating collaborative health management within families. They believed that open access to health data within their family would be beneficial for several reasons. Sharing health data on celiac disease would be educational and informative for our participants about their families’ physical health. Observing others’ health data can help adolescents learn about healthy norms and appropriate practices for managing celiac disease and other issues. Brianna (W4, P12) hoped that her mother would share her health data so that \textit{“I could learn about [celiac disease] because the more we're aware of different issues and diseases… it can help you be more open-minded about everyone's experiences.”} Beyond learning, adolescents were motivated to share health data to engage in their families’ health management and to support loved ones. Julia (W7, P22) explained that she wanted \textit{“to help one another to ensure that we take into consideration everyone's conditions, and also still make meals that everyone can enjoy, and that would benefit the whole health of the family.”} Similarly, Jenny (W4, P11) voiced opportunities to support her family’s health: \textit{“You should be able to sort of understand and learn about what they're going through to continue supporting and helping them throughout their life.”} These comments illustrate that adolescents care deeply about their families and are willing to share health data to understand what their mothers are experiencing and better support them.

\textbf{Actively participating in family health management.} 
Finally, adolescents viewed health AI as an opportunity to involve themselves in family health management. They envisioned the AI systems to be family-oriented and allow direct access for others, including themselves. Julia (W7, P22) commented, \textit{“I think our whole family would end up using [their AI app] if someone in our family had [celiac disease] because now, it’s applicable to everyone.”} Some adolescents saw themselves almost as a health manager, using AI on behalf of their mothers, as Nick (W5, P15) remarked, \textit{“Let’s say you’re grocery shopping instead of your mom, you can use it on her behalf.”} Participants emphasized that they–not just their mothers–should be able to use the health AI tools, highlighting the inclination for direct collaborative use. Similarly, Sarah (W3, P9) described wanting to be a contact person for her mother, in case of emergencies. If her mother accidentally consumed gluten, \textit{“then I can get an alert on my phone and then I can know to go and pick her up and take her home–I have that access to that information,”} underscoring her intent to be more directly involved. These reflections indicate that AI can empower adolescents to not only participate more but also take a more proactive role in their families’ health management.

\subsubsection{AI for Providing Guidance While Respecting Autonomy}
During the activities, our participants welcomed the guidance provided by AI but still wanted the autonomy to make decisions for themselves. Their desire for autonomy was reflected in the findings as they envisioned AI to respect their current lifestyle, give them options for next-steps, and afford opportunities for more complex decision-making. 

\textbf{Embedded into daily routines.} 
When considering AI’s role in everyday life management, adolescents pictured it to be embedded into their daily routines and to respect their own autonomy of living. Opting for an AI app on their phones or small portable devices, they intended the proposed health AI solutions to be functional, effective, and assimilated into their current life. Sarah (W3, P9) described how her family would use their AI tool to \textit{“be able to do things like share food or to enjoy the same meals and have that sense of normalcy.”} Wanting AI to keep everyday routines intact suggests that adolescents want a sense of control amid a family diagnosis and see AI as a way to achieve it. By being available yet unobtrusive, health AI would support autonomy by letting adolescents and their mothers decide if and when to use it. If mothers wanted to continue with their daily tasks, the AI should respect their choice. As Ivy (W6, P16) emphasized, their AI robot would support autonomy as it \textit{“doesn’t have to completely replace your mom’s cooking, it just has to make a better environment that she could do that.”}

\textbf{Offer a variety of options.} 
Similarly, in the health information seeking activities, participants expected AI to support their autonomy by providing a variety of options for next-steps. In the case of personal health assessment, all adolescents took a proactive stance, analyzing their own symptoms with the help of AI (see \textbf{Figure 3}). They then wanted AI to offer a range of suggestions: symptom-checking methods, verified tests for celiac disease and other conditions, and even references to doctors. Having choices builds autonomy by allowing adolescents to have a sense of control and fostering intrinsic motivation in understanding their health status. For instance, Sarah (W3, P9) expressed wanting AI to connect her to \textit{“outside sources,”} such as \textit{“a school therapist,”} or \textit{“scholarly articles,”} or even an \textit{“Instagram page of a Day-in-My-Life of someone with celiac.”} Having suggestions would help adolescents see more variety in the paths they could choose, but they wanted to maintain their independence in making the final choice. Adolescents also wanted AI’s help in obtaining more gluten-free recipes and meal ideas so that they could still enjoy their favorite dishes. Brian (W1, P1) pictured AI helping his mom find restaurants with gluten-free options because \textit{“I’d still want her to enjoy the food that she loves.”} Adolescents wanted more choices, instead of feeling limited by a health condition, and perceived AI to support this. Although many participants expected to explicitly ask AI for advice, it was clear they were leading the investigation, highlighting AI’s ability to reinforce rather than replace their autonomy.

\begin{figure*}[!h]
  \centering
  \includegraphics[width=15.5cm]{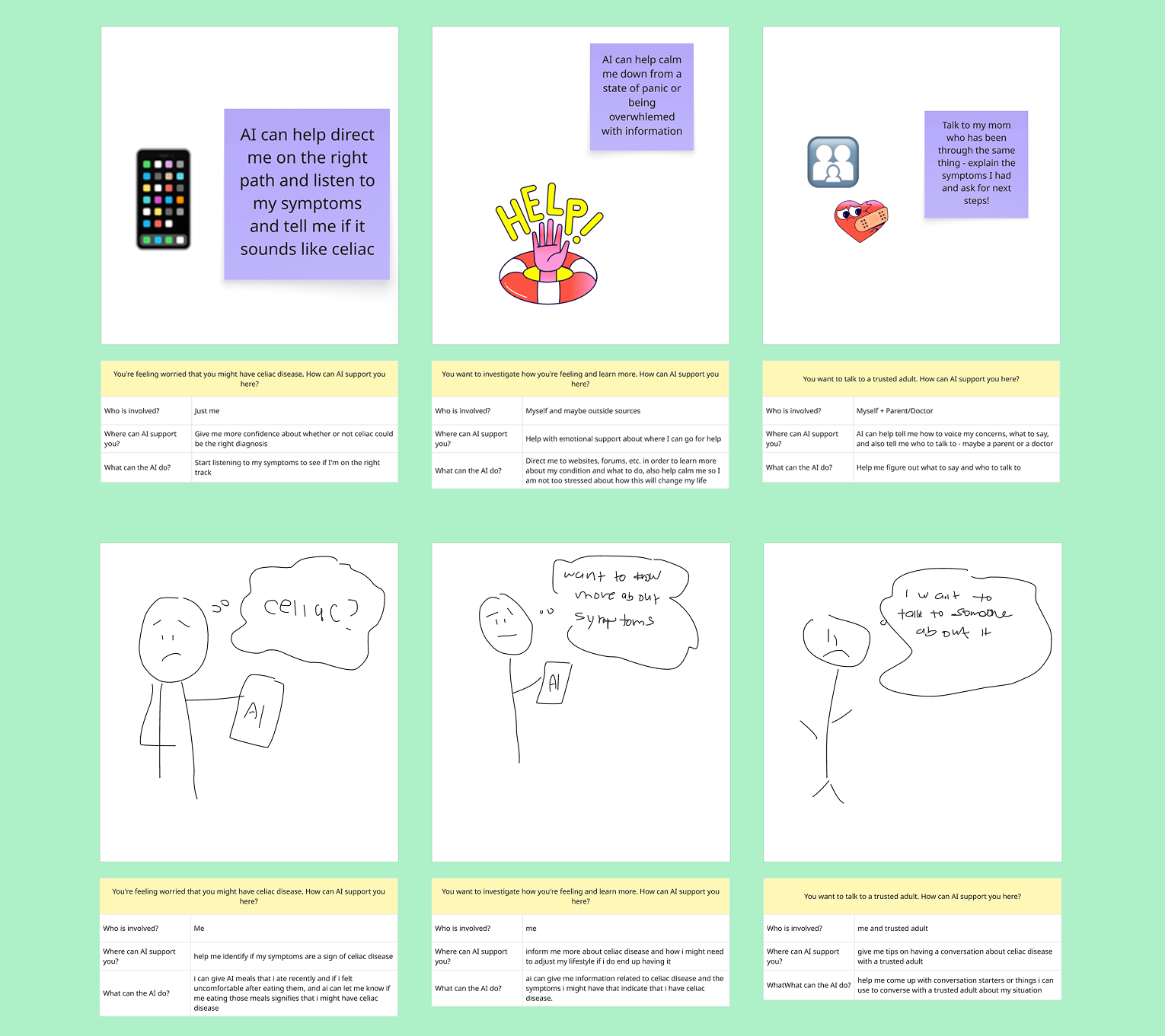}
  \caption{Storyboards from Rae (W3, P8) (top) and Julia (W7, P22) (below)}
  \label{tab:figure3}
  \Description{The image shows two rows of post-it notes with questions.}
\end{figure*}

\textbf{Support complex decision-making.} Our participants also imagined AI affording opportunities for independent, more complex decision-making. Decision-making involves more deliberate thinking and committing to a course of action, such as visiting a doctor for a diagnosis or having a serious discussion with an adult. Throughout the activities, participants were taking initiative with AI as a guide, not a manual to follow. During symptom assessment, Daisy (W1, P2) outlined in her storyboard how she and AI would \textit{“work almost to diagnose diseases, but more what you should do before going to a hospital.”} After evaluation, participants expected AI to continue supporting their investigation by directing them to credible resources–not fully diagnosing them. For instance, adolescents strongly preferred a final diagnosis from a real human doctor, as Jack (W2, P7) emphasized, \textit{“The important part that I want AI to be able to mention to me is that I shouldn’t confirm things by myself… the thing AI should recommend to me is that you should go contact a physician.”} Analyzing one’s symptoms, making sense of a health diagnosis, and deciding to go see a doctor all involve complicated decision-making, and adolescents felt that AI should only guide and refer. When preparing to talk to a trusted adult, adolescents mainly wanted guidance on how to approach the conversation. They were unsure how to initiate difficult conversations with adults and expected to seek advice from AI. Brianna (W4, 12) described how it was challenging at times to engage in a conversation: \textit{“You want to tell a trusted adult… but you don’t know exactly how to bring it up or when’s a right moment to bring it up.”} She and others felt that AI could \textit{“help you come up with ideas on how to approach the conversation.”} In these situations, adolescents highlighted AI’s supportive role and desired opportunities to learn and to make decisions themselves. 

\textbf{Use AI on adolescents’ own terms.} Finally, adolescents envisioned deciding when and how to use AI in their health management journey. They described having AI available for use across different contexts, if and when needed. As mentioned in 4.2.2, participants imagined AI to be embedded and unobtrusive, allowing them to decide when to use it. For example, Ivy (W6, P16) pictured using their AI robot to cook when \textit{“my mom is having a hard day, and because of celiac, she’s feeling down or her energy level isn’t super good.”} They would utilize the AI only when they wanted to. When preparing to talk to a trusted adult, some of our participants could not imagine AI’s role. They felt that AI was helpful enough during their information seeking process, but not necessary for human interactions, such as serious conversations. Instead of using AI to prepare for a conversation, Ashlynn (W3, P10) preferred using it for research first and then she \textit{“would talk to a trusted adult with my little list of information.”} Adolescents viewed AI as a supportive tool that they can use on their own terms–deciding when and how to engage with it.  

\subsection{Perceived Concerns about Trust and Emotional Support from Health AI (RQ2)}
During the study, we found that adolescents had varied attitudes toward trust and emotional support in health AI. Although nuanced, our participants exhibited overall trust in health AI, with several factors affecting this. While they were evenly divided on receiving direct emotional support from AI, the majority found reassurance in the information and the sense of preparedness it provided.

\subsubsection{Nuanced Trust in Health AI}
Our findings reveal that adolescents had considerable trust toward health AI. Although trust varied by participant, overall levels seemed high, with one participant, Samuel (W5, P13), even asserting that he \textit{“fully trusts it.”} This growing trust in AI had three main contributing factors. First, participants reported noticeable improvement in AI recently as Sam (W2, P6) observed, \textit{“I feel like in the past years of using [ChatGPT] really consistently, I feel like it's improved a lot.”} Similarly, Irvin (W7, P23) reflected, \textit{“AI has seen great advancements, technologically… AI before like 2-3 years ago, it couldn’t code, but now, it can make full programs without error.”} Through firsthand use, participants took AI’s improvements as evidence of its growing capabilities. The continual improvement of AI had some adolescents believing that AI could eventually be better than a doctor. Julia (W7, P22) considered the future, \textit{“I think it would have to be a very well trained model for me to say I would trust it over a doctor, but I think eventually it can definitely get there.”} Second, adolescents devised methods of fact-checking the AI, essentially increasing their trust. For instance, when using AI (e.g. ChatGPT), they would double check its responses by searching the Internet themselves or they would ask AI to provide the sources it used. Samuel (W5, P13) explained: \textit{“I sometimes fact check it by, you know, doing research on Google. And you can also ask it for proof, ChatGPT for proof and it'll give you the websites.”} John (W2, P4) detailed a similar thought process: \textit{“I also did ask [ChatGPT] later, what type of resources it used,”} which further enhanced his trust. Their personal methods of verification were sufficient enough to establish strong trust in AI. Finally, adolescents did not perceive celiac disease as highly severe, and because of its prevalence, they had more trust in health AI.

Despite overall trust in health AI, many displayed reservations, revealing nuance. While adolescents appreciated baseline answers to their initial questions, they still exhibited wariness around the validity of the responses for several reasons. To begin, many of our participants did not fully trust AI and preferred to have a human in the loop. They felt the need to confirm the information with a doctor. Although Ivy (W6, P16) thought ChatGPT’s responses were \textit{“relatively accurate,”} she expressed that \textit{“if my mom actually did have celiac disease, I would probably go to an in-person doctor to verify… because everyone else said AI is accurate but only to a certain extent.”} Adolescents felt more confident in health AI when it was certified and endorsed by human experts. Jack (W2, P7) stated his stance, \textit{“But until there’s a trusted interface or a specific AI that the doctors or medical industry themselves use, or like a specific app that’s certified–until then, I would still go to a doctor before confirming my conditions.”} While adolescents felt comfortable trusting AI for low-stakes tasks (e.g., obtaining preliminary information), they did not trust it enough for high-stakes decisions such as diagnosis. Additionally, adolescents explained that they would trust AI more when it provides its sources. When participants mentioned \textit{“sources,”} they were referring to websites, articles, and other content written by respected persons or organizations, and they wanted AI to display them in its system or platform. Brian (W1, P1) noted that ChatGPT \textit{“didn’t give its sources, so I don’t know if the information is all accurate.”} As Julia (W7, P22) thought about future health AI, she expressed, \textit{“I would want [AI] to link research studies that back it up and give me where it’s getting that information,”} which would lead to more confidence. References to external sources seemed to signal to participants that the AI system was built using valid sources. Finally, many adolescents believed that the current state of AI was not advanced enough to be fully trusted. Some of our participants recalled instances where AI (e.g., ChatGPT) was inaccurate. In fact, Sam (W2, P6) was surprised at ChatGPT’s instances of error, \textit{“Sometimes it gives things that are inaccurate, like severely inaccurate. And it's concerning because it feels like it should be way past that by now.”} Yet, many adolescents had confidence that over time and in the future, health AI tools would be more reliable. Irvin (W7, P23) imagined that health AI would drastically improve: \textit{“The AI could possibly give a better assessment, give better advice, lead the mom into a better direction than a doctor can and with even more medical expertise and precision because humans aren’t perfect… and AI isn’t either.”} Our participants anticipated continued AI advancements, with health AI becoming more trustworthy. 
 
Our findings also revealed that some adolescents had misunderstandings of how AI worked, which could ultimately affect their trust. For instance, Brian (W1, P1) stated, \textit{“From what I’ve heard about AI, it looks on the Internet to gather information, right? It doesn’t generate its own information obviously. Well, at least I’d hope not. I hope it’s not making up stuff about restaurants. I hope that it’s checking on Google and on these websites.”} His misunderstanding of LLMs–believing they don’t generate information–could inflate his trust. For example, LLMs can hallucinate when providing source links, a limitation many adolescents in our study did not seem aware of. Sam (W2, P6) also reflected on how ChatGPT could have been helpful years ago to learn more about his grandpa’s rare neural disorder, noting that \textit{“I think with ChatGPT, it would have been really really helpful to get like generic info, and the timeline of the disease… a lot of that was just unknown to us… because it was so rare.”} However, LLMs are less accurate on topics that are rare in their training data \cite{kandpal_large_2023}, which Sam (W2, P6) did not seem to realize. While he genuinely saw the value in AI to support health learning even as a younger teenager, his misunderstanding about how LLMs work could lead to false information, which could be potentially harmful. Although we did not directly ask about how AI works, participants exhibited misconceptions about its functions and limits. 

\subsubsection{Divided on Views of Emotional Support}
Our findings reveal a split in adolescents' views of AI as a source of emotional support. Some participants believed AI could provide direct emotional support as it “showed” empathy and reassurance. Others rejected AI in this role, arguing that AI cannot experience feelings nor genuine empathy nor should it be designed to.

The majority of the participants felt that AI could support them emotionally through information and preparation. Across all the activities, many adolescents reported feeling reassured and emotionally supported when the AI offered details about their symptoms and how to manage celiac disease. When initially searching information using AI, Sarah (W3, P9) noticed how the content about celiac disease was \textit{"daunting"} and \textit{“intimidating”} and she appreciated ChatGPT’s follow-up, \textit{“Some of what it said was really nice. At the end of its response, it was like, ‘Oh I can craft you a list of questions to ask a doctor.’”} She recognized that AI could convey complex or overwhelming information in a comforting way. They also felt that receiving concrete guidance would ease their worries as Ruby (W7, P21) noted, \textit{“I think if I’m feeling worried about [celiac disease], [AI] could help me, maybe give me alternative foods I can eat or something that makes me feel less stressed about having the disease.”} Health AI was imagined to be empathetic through its comforting language and actionable advice as being diagnosed with a health condition was concerning for our participants, even hypothetically for their mothers and themselves. 

Additionally, within their homes, adolescents also believed that their proposed health AI concepts could provide comfort and reassurance not only to their mothers but the whole family as well. For Ashlynn (W3, P10), having a tool that protects her mother from gluten and focuses on prevention would \textit{“reassure the family that the person’s gonna be okay.”} When concerns centered on adolescents' own health (e.g., experiencing symptoms themselves), acquiring more information using the AI was comforting. In his initial moments of worry, Nick (W5, P15) explained that AI could \textit{“support you by answering your questions or concerns or giving you guidance that you might be worried about what’s going on. So AI could come here and help reassure you or guide you on what to do.”} When imagining both pre- and post-diagnosis moments, most adolescents expected to feel comforted by AI’s information. 

Aside from receiving emotional support from better information access and preparation, adolescents were divided in their views of health AI being designed for direct emotional support. They considered AI to be able to explicitly address their feelings in the moment–helping them cope with stress and process difficult emotions. Brian (W1, P1) pictured engaging with AI about \textit{“why my worries might not be valid and it could talk about looking for solutions rather than problems.”} Adolescents believed they could use AI to discuss their feelings and work towards improving their situation, similar to a therapist-style interaction. Sam (W2, P6) reflected on his personal experience using AI for emotional support: \textit{“It does a really really good job with responding in the right way, saying the right things.”} Ashlynn (W3, P10) also shared similar attitudes, noting \textit{“I do think that AI can be a therapist of sorts.”} Beyond informational reassurance, many adolescents perceived AI as capable of directly addressing their distress and improving their emotional wellbeing. However, a few seemed to lean heavily on AI’s empathetic style and sought unequivocal support from AI. Irvin (W7, P23) specifically envisioned AI to support him \textit{“unconditionally”} and to say, \textit{“having celiac is not the end of the world.”} These sentiments illustrate how adolescents recognized and appreciated AI’s compassionate features, yet some might overvalue its simulated “empathy.”

However, others were more cautious and rejected the idea of direct emotional support from AI. Specifically, they maintained that AI could not help them address their distressing emotions because as AI, it is not human and lacks key human qualities, such as true empathy. Ray (W3, P8) explained, \textit{“Well, because [AI’s] not a person, so I feel like it's hard to rely on it emotionally or have it read how you're feeling. So I think it's less reliable in that sense.”} They felt that AI cannot–and should not–be designed to act human, as Daisy (W1, P2) explained, \textit{“Feelings are shared by us [humans], and it shouldn’t be done by AI, just because they won’t grasp the way our mind works.”} Because AI lacks feelings, its ”empathy” came across as inauthentic and superficial as it was just algorithmically generated rather than genuine. They did not want AI to even act human as some found it \textit{“unsettling”} (Daisy W1, P2). Brian (W1, P1) recounted his experience speaking to an AI (e.g., ChatGPT), emphasizing, \textit{“I don't like it when it tries to be like a human. I want it to just be a source of information and stick to the facts.”} Instead of discussing feelings, adolescents believed that AI should only provide information. They felt that technology’s job was to provide information, not emotional support as Daisy (W1, P2) commented, \textit{“I feel like [AI’s] more technology and information based rather than it trying to understand our feelings and how it works.”} While it was acceptable to find comfort in the information that AI provided, it was difficult to imagine being comforted by an AI “acting human.”

\section{DISCUSSION}
In this study, we investigated how adolescents perceive the role of AI in health learning and management within the context of a family celiac disease diagnosis. While previous work has focused on the views of clinicians, adult patients, and parents, our paper sheds light on adolescents’ perceptions of using AI when navigating health situations. 

\subsection{AI's Role in Supporting Adolescent Autonomy in Health Learning and Management}
Our findings revealed that adolescents framed health AI both as a learning tool and a way to participate more actively in their family’s collective health experiences. They expected AI to simplify complex information, support self-assessment, and help them prepare for difficult health-related conversations. Participants further demonstrated initiative by envisioning AI as a learning scaffold–one that could help them gather information about new health conditions, determine actionable next steps, and gather resources to support their mothers. Adolescents also positioned themselves not as passive recipients of information and help, but as active co-users of AI alongside their parents. They positioned AI as a means to participate in their family’s health by advocating for appropriate access to family health data to aid in their mother’s celiac management, reflecting their desire for more independence. Finally, we found that adolescents pictured AI aligning with their daily routines, providing a range of choices, and guiding them through more complex decision-making processes, reflecting their desire for autonomy. Together, these behaviors highlight AI’s potential to foster adolescent autonomy in health management.

Adolescence is a critical developmental period where their biological, cognitive, and socioemotional needs require age-appropriate support, with families serving as the primary context for their transition into adulthood \cite{bonnie_promise_2019, patton_our_2016}. To ensure their overall well-being and long-term health, appropriate health care technologies must be cognizant of adolescents' unique needs \cite{patton_our_2016, frech_healthy_2012}. Prior HCI work indicates that although adolescents desire greater involvement in their health, they require guidance in learning about health and developing their autonomy \cite{zehrung_transitioning_2024, cha_transitioning_2022, hong_care_2016}. We suggest that health AI could enhance adolescents’ learning experiences by employing generative capabilities to streamline, summarize, and explain medical content in a reassuring and developmentally appropriate manner \cite{abreu_enhancing_2024, ayre_new_2024}. Reports of low digital health literacy among adolescents have been documented \cite{taba_adolescents_2022, st_jean_assessing_2017, stellefson_ehealth_2011}, and scholars have found that children and adolescents may have more difficulty in interpreting their health data \cite{su_data-driven_2024, oygur_lived_2021, zehrung_transitioning_2024}. Considering this, the capabilities of health AI offer a critical opportunity to improve digital health literacy among younger populations as they are already using generative AI to summarize or synthesize information for educational purposes \cite{madden_dawn_2024}. 

At the same time, adolescents can also learn about health conditions and management skills through active collaboration with their families. Future health AI should offer flexible settings such as support for multiple users, data-sharing features, and personalized recommendations as a family, which can encourage both family health and adolescent independence \cite{cha_transitioning_2022, su_creating_2024, su_data-driven_2024, zehrung_transitioning_2024, cha_collaborative_2025}. Our participants believed AI can enable family data sharing, which would promote their direct involvement in participating and collaboratively managing family health. Having open access to health data can educate adolescents about what good health means, collaborate as a family on health goals, and prepare them for future health decisions. Beyond simply retrieving facts, these systems can act as learning scaffolds, guiding adolescents to think analytically about health information, to interpret their own health data, and to learn by active engagement and reflection.

Finally, our study underscores the opportunity for health AI to promote adolescents’ autonomy in health learning and management. This aligns with prior research that similarly emphasized the importance of supporting their autonomy and agency in digital environments, particularly regarding data practices, privacy, and technology disengagement \cite{chowdhury_co-designing_2023, wang_12_2023, wang_dont_2022, zhao_i_2019}. We extend this work to the health domain and propose that combining similar strategies with AI capabilities can help adolescents develop the abilities needed to make informed health decisions as they transition into adulthood. However, it is also crucial to consider how the design of AI-powered features (e.g., algorithmic curation or human-like interaction) can influence adolescents’ perceptions, behaviors, and ultimately their health experiences. While current policy frameworks primarily focus on safety (e.g., protecting the child from online harms) \cite{cbe_elizabeth_denham_age_2020, eu_artificial_intelligence_act_future--life-instituteai-act-overview-30-may-2024pdf_2024, world_economic_forum_artificial_2022}, HCI researchers have taken a step further, going beyond to prioritize other themes such as agency \cite{d4cr_d4cr_2022, grace_child-centered_2023, wang_informing_2022, wang_12_2023}. We build on Wang et al.’s call for Child-Centered AI \cite{ge_wang_child-centered_2025}, advocating that health AI systems should be designed with autonomy in mind, actively equipping them with the skills to manage themselves. Reflecting on participant feedback, we recommend that AI systems explicitly frame the outputs as suggestions, alongside mechanisms for young users to better connect with trusted adults and resources.

\subsection{AI's Role in Efficiency for Meaning}
Based on our study, we contend that adolescents perceive the value of AI’s efficiency as a way to reclaim time and energy for meaningful activities, and not as an end in itself. They conceptualized health AI as a means to reduce cognitive burden and support routine tasks in order to create more time for family interactions. Our participants perceived health AI to aid in offloading routine, low-level tasks, while leaving higher-level analysis, complex decision-making, and emotionally significant moments, such as having a conversation with a trusted adult or deciding to visit a doctor, for themselves. Importantly, adolescents drew clear boundaries around what health AI should and should not do, particularly regarding direct emotional support. Participants were divided on the ability of health AI to provide such support, often taking a critical stance. They welcomed emotional reassurance through information but preferred that core activities, such as determining when to seek care, consulting trusted adults, and engaging in family time, remain human-centered. Many felt that human experiences are not easily emulated, and some believed AI should not simulate human-like characteristics at all. Accordingly, AI was viewed as a supplementary tool meant to complement, rather than replace, human experiences. 

Prior work has highlighted AI’s great promise in optimizing efficiency, complementing human abilities, and becoming embedded into users’ daily lives \cite{khalifa_ai_2024, conant_improving_2019}. With its abilities to process, analyze, and organize vast amounts of data, AI has been utilized to offload simpler tasks to increase efficiency and accuracy \cite{lai_human_2019, yang_how_2020}. For the laypersons, recent studies have focused on leveraging AI to find, summarize, and analyze health information, suggesting the possibility for higher search efficiency and accessibility \cite{xiao_powering_2023, ayre_new_2024, yang_talk2care_2024, al_shboul_investigating_2024}. Alongside rapid gains in efficiency, AI’s affective capabilities have also advanced. Recent work has shown that AI can also support more emotional aspects of human experiences \cite{jung_ive_2025, lee_i_2020, sweeney_can_2021, inkster_empathy-driven_2018}. In particular, designing human characteristics such as empathy may further aid in emotional regulation, improve accessibility, and reduce symptoms of depression and anxiety \cite{fitzpatrick_delivering_2017, jung_ive_2025, daher_empathic_2020, you_beyond_2023}. However, our findings reveal that adolescents interpret efficiency and affective design differently. First, while our participants valued saving time on low-effort tasks, they recognized that saved time as an opportunity to engage in more purposeful activities. Second, although they acknowledged a place for empathy in AI, adolescents in our study resisted AI that \textit{“acts human.”} While human-like design can convey warmth and empathy to the users, overly empathetic and agreeable AI can be risky, especially if it substitutes for human support. Recent work has highlighted the notable harms that can come from engaging with AI socially and emotionally, such as emotional dependence \cite{laestadius_too_2024}, harassment, or even violence \cite{zhang_dark_2025}. By being too emotionally supportive, adolescents could rely on AI rather than seeking human support, and they could experience very serious harms from AI. At the most troubling extreme, adolescents could experience severe mental health issues and resort to AI instead of seeking professional help. Recent news has reported on parents alleging that AI chatbots enabled their children to commit suicide \cite{clare_duffy_there_2024, angela_yang_family_2025}, underscoring the significance of this issue. 

Bringing these strands of work together, we argue that AI enables efficiency in meaning, or supports low-level tasks to make time for intrinsically valuable activities. This ties into the concept of ‘dialectical activities’ \cite{zhang_searching_2024}, or human activities whose values are intrinsic and emerge only through repeated engagement, such as parenting, learning, or conducting research. Rather than viewing AI as replacing these dialectical activities, adolescents saw it as a guiding tool that complements and respects their lives. Although AI’s efficiency in health learning and disease management is valued, we should look towards affording meaningful experiences and value-sensitive approaches, which can build independence, self-control, and overall well-being. To protect and uplift adolescents, we suggest that future health AI should de-emphasize certain human-like designs, promote reflection through engagement, and enable meaningful, dialectical activities.

\subsection{Calibrating Adolescents' Qualified Trust in Health AI}
In our study, adolescents exhibited qualified trust in health AI–a calibrated and conditional form of trust. Their trust was contextually grounded, developing through consistent system performance while maintaining awareness of its limitations. It was further shaped by factors such as health activity and AI misconceptions. Adolescents placed greater trust in AI for information seeking, everyday routine tasks, and celiac management, even anticipating deeper integration into their lifestyles. However, they preferred to have a human involved for high-risk situations: confirming health information, making a final diagnosis, and confiding in them about their personal health journeys. Some adolescents also had misconceptions about AI, which could miscalibrate their trust. 

Prior work has noted that the severity of the health condition could affect trust in health AI \cite{lee_understanding_2025}, but our findings reveal more nuance. Although participants viewed celiac disease as relatively low in severity, they exerted different levels of trust, depending on AI’s function. Clinically, they did not trust AI to diagnose celiac disease, but to assist with information seeking and preparation for a medical visit. Practically, they did not trust AI to fully arrange their lives, but to make existing routines easier. Emotionally, they did not trust AI to provide therapeutic support but to communicate information in a comforting tone to release the stress of navigating a new health condition. These findings highlight the complexity of their trust toward health AI, which is multi-dimensional, context-dependent, and evolves with experience. 

Our study suggests more work is needed to deeply examine adolescents’ trust in health AI. Given that trust is complex and evolves over time, there is a need for more comprehensive research that considers not only their perception but also their varied health contexts. Prior work has explored how context and perceived risk can affect trust \cite{esmaeilzadeh_patients_2021, kim_how_2024, bach_systematic_2024}. Building on prior work, this study uncovered that adolescents’ trust varied depending on the function of AI and the activities it supports, suggesting that further work is needed to investigate the differences in adolescents’ interactions and trust with health AI across varying health activities. 
Additionally, some adolescents revealed misunderstandings about AI, such as the assumption that LLMs did not produce new, original text but instead served as a search engine. This misconception of how AI functions is associated with numerous risks, including adolescents’ miscalibration of trust. For instance, AI can produce inaccurate health advice, and acting on such advice can be harmful \cite{singhal_towards_2023, agarwal_medhalu_2024, kim_medical_2025}. Without understanding that AI can produce fabricated information or biased content, adolescents, who may more readily accept seemingly credible outputs as valid, could take AI’s guidance too seriously, leading to adverse consequences.

To combat miscalibration and ensure qualified trust in health AI, we advocate for supporting adolescents’ AI literacy within the health context. AI literacy is the ability to critically understand, evaluate, and use AI technologies \cite{long_what_2020, mills_AI_2024}, and prior work has shown that AI literacy is crucial for users to assess and calibrate their trust. HCI studies have explored how to develop AI literacy for adolescents \cite{charisi_empowering_2020, wang_informing_2022, sharma_robot_2025}, which is particularly crucial as younger people are more likely to use AI for health information \cite{yun_online_2025}. Improving AI literacy offers substantial benefits. First, understanding how these systems work can increase one’s ability to use AI applications more effectively and safely \cite{kimiafar_artificial_2023, kong_evaluation_2021}. Second, understanding algorithmic systems can empower the youth to develop new perspectives on AI models and identify ways to improve them, underscoring the benefits of critical AI literacy \cite{morales-navarro_youth_2024, noh_youth_2025}. Finally, broad digital literacy is empowering, as it “provides people with new abilities and ways to participate in a digital society” \cite{makinen_digital_2006}. This can equip adolescents with the capability to effectively use AI and foster a stronger sense of control over their lives. To achieve this, we suggest that researchers, educators, and designers work together with adolescents to create more adolescent-friendly educational resources on AI literacy. Placing the youth as advisors in participatory design research has shown to build trust, foster deep engagement, and truly integrate adolescent perspectives and needs into the design process \cite{noh_youth_2025}. 

Taken together, our study points to growing yet qualified trust in health AI among adolescents. At the same time, misconceptions about how these AI systems generate outputs can miscalculate trust and lead to potentially harmful decisions, underscoring the need for health AI literacy. This lens opens opportunities to provide targeted educational support and empower adolescents to use AI effectively and safely. Future work should also explore more deeply how AI can assist adolescents in varying health contexts and during health-related activities while properly calibrating their trust.  

\section{CONCLUSION}
To understand adolescents' envisioned roles of AI in supporting their own and their family’s health management, we conducted 7 workshops with 23 adolescents to explore their perceptions of health AI. Our findings show that adolescents had four main expectations: enhance health understanding and help-seeking, reduce cognitive burden, mediate family collaborative health management, and provide guidance while respecting autonomy. Participants also voiced concerns regarding trust and receiving emotional support from health AI. These findings suggest that adolescents envision AI’s role to promote their health autonomy as they mature into adulthood. They also perceived AI as a tool that transforms efficiency into meaning, using its capabilities to take over routine tasks and create space for meaningful activities. We suggest that future health AI systems be designed to foster adolescent autonomy, promote health AI literacy, and enable time for meaningful, dialectical activities. With this paper, we intend to inform and inspire further exploration of adolescents' unique needs and perceptions regarding health AI. 

\begin{acks}
We thank our participants for their time and engagement. We also thank the summer programs hosted by UCI (MedAcademy, ICS Summer Academy, COSMO, and Summer Surgery Program) for their support in study recruitment. This study was supported by grants from the National Science Foundation (NSF) grant number 2211923 and GIM Funding from UCI School of Medicine. 
\end{acks}

\bibliographystyle{ACM-Reference-Format}
\bibliography{references}

\end{document}